\documentclass[a4paper,11pt]{article}
\usepackage{amsbsy}
\usepackage{amsfonts}
\usepackage{amsmath}
\usepackage{amssymb}
\usepackage[titletoc,toc,title]{appendix}
\usepackage{authblk}
\usepackage{braket}
\usepackage{cancel}
\usepackage{caption}
\usepackage{cite}
\usepackage{comment}
\usepackage{enumitem}
\usepackage{epsfig}
\usepackage{etoolbox}
\usepackage{float}
\usepackage[margin=2.5cm]{geometry}
\usepackage{graphicx}
\usepackage{setspace}
\usepackage{hyperref}
\usepackage{physics}
\usepackage{titlesec}

\makeatletter
\patchcmd{\@maketitle}{center}{flushleft}{}{}
\patchcmd{\@maketitle}{center}{flushleft}{}{}
\patchcmd{\@maketitle}{\LARGE}{\LARGE\sffamily\bfseries}{}{}
\makeatother

\makeatletter
\def\maketitle{{%

\AB@maketitle}}

\renewcommand\AB@affilsepx{\protect\\\vskip1em\protect\Affilfont}

\makeatother

\renewcommand\Affilfont{\normalsize\normalfont\itshape\small}

{\makeatletter\g@addto@macro\bfseries{\boldmath}\makeatother}

\makeatletter
\g@addto@macro\ps@plain{%
\def\@oddfoot{\reset@font\hfill-- \thepage\ --\hfill}%
\let\@evenfoot\@oddfoot
}
\makeatother
\titleformat*{\section}{\normalfont\large\bfseries}
\titleformat*{\subsection}{\normalfont\normalsize\bfseries}
\titleformat*{\subsubsection}{\normalfont\normalsize\bfseries}
\titleformat*{\paragraph}{\normalfont\normalsize\bfseries}
\titleformat*{\subparagraph}{\normalfont\normalsize\bfseries}

\hypersetup{%
citecolor=red,
colorlinks=true,
filecolor=red,
linkcolor=blue,
linktocpage=true,
urlcolor=blue
}

\begin{document}

\title{Holographic entanglement negativity for a single subsystem in conformal field theories with a conserved charge
\vskip23pt plus0.06fil minus13pt
\vskip -5pt
\hrule height 1.5pt
\vskip23pt plus0.06fil minus13pt
\vskip -25pt
}

\author[1]{Sayid Mondal\thanks{\noindent E-mail:~ {\tt sayid.mondal@gmail.com}}}
\author[2]{Boudhayan Paul\thanks{\noindent E-mail:~ {\tt paul@iitk.ac.in}}}
\author[2]{Gautam Sengupta\thanks{\noindent E-mail:~ {\tt sengupta@iitk.ac.in}}}
\author[3]{Punit Sharma\thanks{\noindent E-mail:~ {\tt punit-sharma-1@uiowa.edu}}}

\affil[1]{%
\parbox[t]{\linewidth}{%
Center for High Energy Physics and Department of Physics\\

Chung-Yuan Christian University, Chung-Li 320, Taiwan
}
}

\affil[2]{%
\parbox[t]{\linewidth}{%
Department of Physics\\

Indian Institute of Technology\\ 

Kanpur 208016, India
}
}
\affil[3]{%
\parbox[t]{\linewidth}{%
Department of Physics and Astronomy\\

The University of Iowa\\ 

Iowa City, IA 52242, USA
}
}

\date{}

\begin{NoHyper}
\maketitle
\end{NoHyper}

\thispagestyle{empty}

\begin{abstract}

We utilize a holographic construction to compute the entanglement negativity for bipartite mixed state configurations of a single subsystem in $CFT_d$s with a conserved charge dual to bulk $AdS_{d+1}$ geometries. In this context, we obtain the holographic entanglement negativity for single subsystems with long rectangular strip geometries in $CFT_d$s with a conserved charge dual to bulk non extremal and extremal Reissner-Nordstr\"{o}m (RN)-$AdS_{d+1}$ black holes at the leading order in a perturbation theory. We demonstrate that the holographic entanglement negativity computed involves the elimination of thermal contributions at the leading order confirming earlier results in the literature. This also conforms to quantum information theory expectations and constitutes further consistency checks for the holographic construction.

\end{abstract}

\clearpage
\hrule
\tableofcontents
\bigskip\medskip
\hrule
\bigskip\bigskip
\clearpage

\setlength\parindent{1.2\parindent}
\setstretch{1.1}

\section{Introduction}
\label{sintro}

The recent past has witnessed the emergence of quantum entanglement as a central theme in the study of many body systems relevant to diverse fields from condensed matter physics to quantum gravity and black holes. For bipartite pure states the {\it entanglement entropy} serves as an appropriate entanglement measure. Remarkably for $(1+1)$-dimensional conformal field theories ($CFT_{1+1}$s) an explicit computation for the entanglement entropy utilizing the replica technique was established in \cite{Calabrese:2004eu,Calabrese:2009ez,Calabrese:2009qy,2009JPhA...42X0301C,Calabrese:2010he}. For bipartite mixed states however the entanglement entropy incorporates irrelevant correlations and is hence unsuitable for characterization of the entanglement. In quantum information theory, several measures have been proposed for the characterization of mixed state entanglement which are in general not easily computable. Vidal and Werner in a significant work \cite{Vidal:2002zz} proposed a computable measure termed {\it entanglement negativity (logarithmic negativity)} for bipartite mixed state entanglement which characterizes an upper bound on the distillable entanglement for the mixed state.\footnote{Distillable entanglement is the amount of pure entanglement that can be extracted from a given state by means of local operations and classical communication (LOCC).} This measure was defined as the logarithm of the {\it trace norm} for the partially transposed reduced density matrix with respect to one of the subsystems and shown to be an entanglement monotone \cite{Plenio:2005cwa}.\footnote{For an excellent review see \cite{Rangamani:2014ywa}.} Interestingly in a series of articles \cite{Calabrese:2012ew,Calabrese:2012nk,Calabrese:2014yza} the entanglement negativity for various bipartite pure and mixed states in a $CFT_{1+1}$ could be obtained through an appropriate replica technique.

In a dif{}ferent context a holographic characterization of the entanglement entropy for $d$-dimensional conformal field theories ($CFT_d$s) dual to bulk $AdS_{d+1}$ geometries was conjectured in a seminal work by Ryu and Takayanagi (RT) \cite{Ryu:2006bv,Ryu:2006ef}. Their proposal described the holographic entanglement entropy of a subsystem in the dual $CFT_d$ in terms of the area of a bulk codimension two static minimal surface homologous to the subsystem (RT Surface). The RT conjecture and its covariant generalization by Hubeny, Rangamani and Takayanagi (HRT) \cite{Hubeny:2007xt} were proved in a series of subsequent works \cite{Fursaev:2006ih,Headrick:2019eth,Casini:2011kv,Faulkner:2013yia,Lewkowycz:2013nqa, Dong:2016hjy}. The RT/HRT conjectures inspired the remarkable development of the field of holographic quantum entanglement in dual $CFT_d$s summarized in \cite{Nishioka:2009un,Takayanagi:2012kg,Fischler:2012ca,Fischler:2012uv,Cadoni:2010kla,Chaturvedi:2016kbk} and references therein.\footnote{Note that the field is extremely diverse and the list of references is not meant to be exhaustive.}

The above developments naturally brought the critical issue of a holographic characterization for the entanglement negativity in dual $CFT_d$s into sharp focus. This significant issue was initially partially addressed for pure states in \cite{Rangamani:2014ywa} although a clear holographic prescription for generic bipartite mixed states in dual $CFT_d$s remained an outstanding issue. In an interesting communication the authors in \cite{Chaturvedi:2016rcn} proposed a holographic entanglement negativity conjecture for the bipartite states of a single interval in $CFT_{1+1}$s dual to bulk pure $AdS_3$ geometries and Euclidean BTZ black holes. This proposal involved a certain algebraic sum of the lengths of bulk geodesics homologous to appropriate combinations of the single interval and other auxiliary intervals. Remarkably the results for the holographic entanglement negativity computed from their proposal were in exact agreement with the corresponding $CFT_{1+1}$ replica technique results in the large central charge limit described in \cite{Calabrese:2012ew,Calabrese:2012nk,Calabrese:2014yza}. This construction was further substantiated through a rigorous large central charge analysis utilizing the monodromy technique in \cite{Malvimat:2017yaj} for the $AdS_3/CFT_2$ scenario. Subsequently in a series of articles this holographic construction and its covariant generalization were utilized to obtain the entanglement negativity for diverse mixed state configurations in the $AdS_3/CFT_2$ scenario \cite{Chaturvedi:2016opa,Chaturvedi:2016rcn,Jain:2017aqk,Jain:2017uhe,Malvimat:2018ood,Malvimat:2018txq}.

A higher dimensional generalization of the above holographic conjecture for generic $AdS_{d+1}/CFT_d$ scenarios involving specific algebraic sums of the areas of the RT surfaces for certain combinations of subsystems, was subsequently proposed and utilized to compute the entanglement negativity of bipartite states of subsystems with long rectangular strip geometries in dual $CFT_d$s \cite{Chaturvedi:2016rft,Jain:2017xsu,Jain:2018bai,KumarBasak:2020viv}. The results conformed to quantum information theory expectations and reproduced certain significant features for the entanglement negativity also observed for the $AdS_3/CFT_2$ scenario which served as crucial consistency checks. Very recently a plausible bulk proof of these holographic entanglement negativity conjectures from a gravitational path integral perspective, along the lines of \cite{Faulkner:2013yia,Lewkowycz:2013nqa} for spherical subsystem geometries based on \cite{Dong:2021clv} was described in \cite{KumarBasak:2020ams}.\footnote{A review of this plausible proof for single subsystems with spherical geometries has been reviewed in appendix \ref{ahendrv}.} However a generalization of this for generic subsystem geometries is still a non trivial outstanding issue.

We should mention here that subsequent to the developments described above, the authors in \cite{Kudler-Flam:2018qjo,Kusuki:2019zsp} proposed an alternate holographic entanglement negativity conjecture for bipartite states in dual $CFT_d$s involving the minimal entanglement wedge cross section (EWCS) \cite{Takayanagi:2017knl} backreacted by the cosmic brane for the conical defect of the bulk replicated geometry. For spherical entangling surfaces this backreaction could be computed and resulted in a dimension dependent constant numerical factor for the entanglement negativity. For the $AdS_3/CFT_2$ scenario the holographic results from their proposal matched with the corresponding field theory replica technique results in the large central charge limit modulo certain constants arising from the Markov gap \cite{Hayden:2021gno}. The above holographic construction was further refined in \cite{KumarBasak:2020eia} to resolve an outstanding issue and demonstrated the equivalence of the two proposals modulo the constants mentioned above.\footnote{Note that the constant backreaction factor $\chi_d=\frac{3}{2}$ for the $AdS_3/CFT_2$ scenario was incorporated in the proposal based on the algebraic sum of the lengths of bulk geodesics homologous to appropriate combinations of intervals described in \cite{Chaturvedi:2016rcn}.} However this alternative proposal requires further substantiation for the higher dimensional $AdS_{d+1}/CFT_d$ scenario.\footnote{See also the discussions in \cite{Jokela:2019ebz}.}

In this article we extend the higher dimensional holographic construction to compute the entanglement negativity for bipartite states of single subsystems described in \cite{Chaturvedi:2016rcn} to $CFT_d$s with a conserved charge dual to bulk non extremal and extremal $AdS_{d+1}$-RN black holes. Our results provide further substantiation for the holographic conjecture for the entanglement negativity of bipartite states in higher dimensional $CFT_d$s. In this context following \cite{Chaturvedi:2016rcn} we also consider subsystems with long rectangular strip geometries and utilize a perturbative technique described in \cite{Fischler:2012uv,Fischler:2012ca,Kundu:2016dyk,Kundu:2016cgh}, to compute the areas of RT surfaces involving non trivial limits of certain parameters relevant for the bulk $AdS_{d+1}$-RN geometry.\footnote{\label{nmeijer} Note that an exact closed form analytic evaluation of the area has been performed through the use of Meijer G-functions in \cite{Erdmenger:2017pfh}. However the leading order behavior necessary for our purpose is correctly captured through the perturbative techniques utilized here.} Interestingly as described in the literature, through our construction we reproduce certain significant features (such as elimination of the thermal contributions) of the holographic entanglement negativity for bipartite states in $CFT_2$s and $CFT_d$s described in \cite{Chaturvedi:2016rcn,Chaturvedi:2016rft}. This conforms to quantum information theory expectations of the negativity being an upper bound on the distillable entanglement.

Note that in the context of the developments described in \cite{Kudler-Flam:2018qjo,Kusuki:2019zsp,KumarBasak:2020eia} our results also involve backreaction ef{}fects due to the cosmic brane arising from the conical defect of the replicated bulk geometry in the replica limit. However the explicit determination of these ef{}fects remains a non trivial outstanding issue for arbitrary subsystem geometries, in particular the long rectangular strip geometries investigated in this article.

In this connection we state here that in \cite{Belin:2013uta} the authors provided another construction to characterize the modification of the RT surfaces due to the presence of a charged bulk geometry. In this article they considered a charged topological black hole with a hyperbolic horizon coupled to a dual Maxwell gauge field to compute charged holographic R{\'e}nyi entropies and the holographic entanglement entropy for spherical entangling surfaces in the dual $CFT_d$. Interestingly the corresponding charged R{\'e}nyi entropies and the entanglement entropy for such spherical domains in the $CFT_d$ could be computed using a generalized replica technique for a thermal path integral via a background gauge field with a non trivial Wilson line on the Euclidean time circle following an appropriate conformal transformation.

This article is organized as follows. In section \ref{shenr} we describe the holographic entanglement negativity construction for the bipartite configuration of a single subsystem characterized by long rectangular strip geometry in $CFT_d$s dual to bulk $AdS_{d+1}$ geometries. Subsequently in section \ref{shenhd} the construction is employed to compute the holographic entanglement negativity for mixed state configuration of a single subsystem in $CFT_d$s with a conserved charge dual to bulk non extremal and extremal $AdS_{d+1}$-RN black holes. Finally in section \ref{ssum} we summarize our results and present our conclusions. We also provide some involved and lengthy results in the appendices.

\section{Holographic entanglement negativity}
\label{shenr}

In this section we present a brief review of the holographic entanglement negativity conjecture for the bipartite state of a single subsystem in a $CFT_d$ at a finite temperature dual to bulk $AdS_{d+1}$ geometries as described in \cite{Calabrese:2012ew,Calabrese:2012nk,Calabrese:2014yza,Chaturvedi:2016rft,Chaturvedi:2016rcn}. In this context we begin with recapitulating the corresponding issue in the $AdS_3/CFT_2$ scenario \cite{Chaturvedi:2016rcn}. For this purpose we consider a bipartite system ($A\cup A^c$) which is described by an interval $A$ of length $l$ and its complement $A^c$ in the dual $CFT_{1+1}$. To describe this construction it is also required to consider two auxiliary large but finite intervals $B_1$ and $B_2$ each of length $b$ on either side of $A$ and adjacent to it such that $B=(B_1\cup B_2)\subset A^c$.
The intervals may be represented as $B_1=(-b-l/2,-l/2), A=(-l/2,l/2),B_2=(l/2,b+l/2)$.
The entanglement negativity for the mixed state of the single interval at a finite temperature in the dual $CFT_{1+1}$ may then be defined through a replica technique as follows \cite{Calabrese:2012ew}
\begin{equation}
\label{eendef}
\mathcal{E}=\lim_{B\to A^c}
\lim_{n_e\to 1}\ln\mathrm{Tr}\left(\rho_{AB}^{T_B}\right)^{n_e},
\end{equation}
where $\rho_{AB}^{T_B}$ is the partial transpose of the reduced density matrix $\rho_{AB}$ with respect to the subsystem $B$,
and the replica limit $n_e\to 1$ is an analytic continuation of even sequences of $n_e$ to $n_e=1$. For two arbitrary subsystems $X$ and $Y$, the partial transpose $\rho_{XY}^{T_Y}$ of the reduced density matrix $\rho_{XY}$, with respect to the subsystem $Y$, may be defined as
\begin{equation}
\label{pt_def}
\mel{e^{(X)}_ie^{(Y)}_j}{\rho_{XY}^{T_Y}}{e^{(X)}_ke^{(Y)}_l}=\mel{e^{(X)}_ie^{(Y)}_l}{\rho_{XY}}{e^{(X)}_ke^{(Y)}_j},
\end{equation}
where ${|e^{(X)}_i\rangle}$ and $|e^{(Y)}_j\rangle$ describe orthonormal bases for the Hilbert spaces $\mathcal{H}_X$ and $\mathcal{H}_Y$ respectively.
Note that the quantity $\mathrm{Tr}(\rho_{AB}^{T_B})^{n_e}$ may be expressed as a twist correlator in the $CFT_{1+1}$ appropriate to the bipartite configuration. For the case of the bipartite state of a single interval in the dual $CFT_{1+1}$ this is given by a four point twist field correlator on an infinitely long cylinder and the corresponding entanglement negativity is given as follows \cite{Calabrese:2014yza}
\begin{equation}
\label{eensg}
\mathcal{E}=\lim_{l/b\to 0}~\lim_{n_e\to 1}\ln\left[\left\langle\mathcal{T}_{n_e}(-b-l/2)\overline{\mathcal{T}}^2_{n_e}(-l/2)\mathcal{T}^2_{n_e}(l/2)\overline{\mathcal{T}}_{n_e}(b+l/2)\right\rangle_{cyl}\right],
\end{equation}
where $l/b\to 0$ is the bipartite limit $B=(B_1\cup B_2)\to A^c$. It is important to note that the order of the limits in eq.\@ \eqref{eensg} is crucial where the bipartite limit ($l/b\to 0$) should be taken only after the replica limit $(n_e\to 1)$.

The twist fields $\mathcal{T}$s, $\overline{\mathcal{T}}$s above provide the boundary conditions between the $n_e$ replicas and reduce to primary fields on the complex plane in the replica limit $ n_e\to 1$. As described in \cite{Malvimat:2017yaj} the universal part of the four point twist field correlator described above may be obtained in the large central charge limit through a rigorous monodromy analysis as follows 
\begin{equation}
\label{etwistlc}
\left\langle\mathcal{T}_{n_e}(z_1)\overline{\mathcal{T}}^2_{n_e}(z_2)\mathcal{T}^2_{n_e}(z_3)\overline{\mathcal{T}}_{n_e}(z_4)\right\rangle_\mathbb{C}\approx\frac{1}{x^{\Delta^{(2)}_{n_e}}z_{23}^{2\Delta^{(2)}_{n_e}}}\,,
\end{equation}
where the cross ratio $x\equiv\frac{z_{12}z_{34}}{z_{13}z_{24}}$ and $z_{ij}=z_i-z_j$ and the scaling dimensions $\Delta_{n_e}$ and $\Delta_{n_e}^{(2)}$ of the operator $\mathcal{T}_{n_e}$ and $\mathcal{T}_{n_e}^2$ are related to each other as follows
\begin{equation}
\label{etwistdim}
\Delta_{n_e}=\frac{c}{12}\left(n_e-\frac{1}{n_e}\right),\qquad\qquad\Delta_{n_e}^{(2)}=2\Delta_{\frac{n_e}{2}}=\frac{c}{6}\left(\frac{n_e}{2}-\frac{2}{n_e}\right).
\end{equation}
From the two point correlator of the twist fields in a $CFT_{1+1}$, given as
\begin{equation}
\label{e2ptfn}
\left\langle\mathcal{T}_{n_e}(z_i)\overline{\mathcal{T}}_{n_e}(z_j)\right\rangle_\mathbb{C}~\sim~\left\lvert z_{ij}\right\rvert^{-2\Delta_{\mathcal{T}_{n_e}}},
\end{equation}
it is observed that the large central charge form of the four point twist correlator in eq.\@ \eqref{etwistlc} may be expressed in terms of the products of certain two point twist field correlators. Using the $AdS_3/CFT_2$ dictionary these two point twist correlators may in turn be expressed in terms of the lengths of space like geodesics in the bulk $AdS_3$ geometry (with length scale $R$) homologous to certain combinations of the intervals as follows
\begin{equation}
\left\langle\mathcal{T}_{n_e}(z_i)\overline{\mathcal{T}}_{n_e}(z_j)\right\rangle_\mathbb{C}~\sim~e^{-\frac{\Delta_{n_e}\mathcal{L}_{ij}}{R}}\,,\nonumber
\end{equation}
\begin{equation}
\label{e2ptfnblk}
\left\langle\mathcal{T}_{\frac{n_e}{2}}(z_k)\overline{\mathcal{T}}_{\frac{n_e}{2}}(z_l)\right\rangle_\mathbb{C}~\sim~e^{-\frac{\Delta_{\frac{n_e}{2}}\mathcal{L}_{kl}}{R}}\,.
\end{equation}
In the above expressions $\mathcal{L}_{ij}$ is the length of the geodesic homologous to the interval described by the endpoints $z_i$ and $z_j$. From eqs.\@ \eqref{eensg}, \eqref{etwistlc} and \eqref{e2ptfnblk}, the holographic entanglement negativity may then be expressed as
\begin{equation}
\label{ehenlen}
\mathcal{E}=\lim_{B\to A^c}~\left.\frac{3}{16G_N^{(3)}}\middle(2\mathcal{L}_A+\mathcal{L}_{B_1}+\mathcal{L}_{B_2}-\mathcal{L}_{A\cup B_1}-\mathcal{L}_{A\cup B_2}\right),
\end{equation}
where we have utilized the Brown-Henneaux formula $c=\frac{3R}{2G_N^{(3)}}$ \cite{Brown:1986nw}. As discussed in the Introduction, the expression for the holographic entanglement negativity above contains an overall constant numerical factor due to the backreaction of the cosmic brane for the bulk conical defect in the replica limit.\footnote{For the $AdS_3/CFT_2$ scenario where the entangling surfaces are just points this overall dimension dependent backreaction factor $\chi_d$ may be explicitly determined as $\chi_2=\frac{3}{2}$ as described in \cite{Nishioka:2018khk,Kudler-Flam:2018qjo}.} Note that using the RT prescription for the $AdS_3/CFT_2$ scenario the above equation may be expressed as follows
\begin{equation}
\label{ehenhee}
\mathcal{E}=\lim_{B\to A^c}\left.\frac{3}{4}\middle(2 S_A+S_{B_1}+S_{B_2}-S_{A\cup B_1}-S_{A\cup B_2}\right).
\end{equation}

Given the $AdS_3/CFT_2$ scenario discussed above, it was conjectured in \cite{Chaturvedi:2016rft} that in the context of the $AdS_{d+1}/CFT_d$ framework the corresponding holographic entanglement negativity for the mixed state of a single subsystem $A$ may be expressed in the bipartite limit $(B\to A^c)$ as
\begin{equation}
\label{ehensg}
\mathcal{E}_A=\lim_{B\to A^c}\left.\frac{3}{16G_N^{(d+1)}}\middle(2\mathcal{A}_A+\mathcal{A}_{B_1}+\mathcal{A}_{B_2}-\mathcal{A}_{A\cup B_1}-\mathcal{A}_{A\cup B_2}\right),
\end{equation}
where $\mathcal{A}_\gamma$ denotes the area of the RT surface homologous to the subsystem $\gamma$, and $G_N^{(d+1)}$ is the gravitational constant in $(d+1)$ dimensions.\footnote{As discussed in the Introduction a plausible proof of the holographic conjectures expressed in eqs.\@ \eqref{ehenlen} and \eqref{ehensg} for spherical entangling surfaces discussed in \cite{KumarBasak:2020ams} from replica symmetry breaking saddles for the bulk gravitational path integral \cite{Dong:2021clv} has been described in appendix \ref{ahendrv}.}

For the case when the subsystems $B_1$ and $B_2$ are symmetric with respect to the subsystem $A$, eq.\@ \eqref{ehensg} simplifies to
\begin{equation}
\label{ehensgsym}
\mathcal{E}_A=\lim_{B\to A^c}\left.\frac{3}{8G_N^{(d+1)}}\middle(\mathcal{A}_A+\mathcal{A}_{B_1}-\mathcal{A}_{A\cup B_1}\right).
\end{equation}
Using the standard Ryu-Takayanagi (RT) formula given as
\begin{equation}
\label{eheesg}
\mathcal{S}_A=\frac{\mathcal{A}_A}{4G_N^{(d+1)}}\;,
\end{equation}
the above conjecture for the holographic entanglement negativity for the bipartite state of a single subsystem $A$ may be expressed as%
\footnote{Note that this is a purely holographic result valid for the corresponding large $c$ limit of the dual $CFT$. This is explicit for the $AdS_3/CFT_2$ scenario where exact computations are possible for both the bulk and the boundary. To check its validity for the full quantum information theory measure, it would be required to compute the non universal contributions to the $CFT_2$ correlation functions, which is extremely non trivial.}
\begin{equation}
\label{ehenheesym}
\mathcal{E}_A=\lim_{B\to A^c}\left.\frac{3}{2}\middle(S_A+S_{B_1}-S_{A\cup B_1}\right).
\end{equation}

Note that as discussed earlier for the $AdS_3/CFT_2$ scenario, the expressions in eqs.\@ \eqref{ehenhee} and \eqref{ehensg} for the $AdS_{d+1}/CFT_d$ framework will also involve ef{}fects due to the backreaction of the cosmic brane for the conical defect of the replicated bulk geometry. For arbitrary subsystem geometries the determination of the explicit form of these ef{}fects is still a non trivial open issue in particular for the long rectangular strip geometries considered in this article. However for spherical entangling surfaces the explicit form of the backreaction factor is described in \cite{Nishioka:2018khk,Kudler-Flam:2018qjo}.

\section{Holographic entanglement negativity for $CFT_d$ dual to $AdS_{d+1}$-RN}
\label{shenhd}

We now utilize the construction reviewed above to compute the holographic entanglement negativity for bipartite states of single subsystems in $CFT_d$s with a conserved charge, dual to bulk $AdS_{d+1}$-RN geometries. We employ a perturbative expansion of the areas of the corresponding bulk RT surfaces, in terms of the temperature $T$ and the chemical potential $\mu $ conjugate to the charge $Q$ of the $AdS_{d+1}$-RN black hole. It is convenient to express the holographic entanglement negativity in terms of the total energy $\varepsilon$ of the dual $CFT_d$, and an {\it ef{}fective temperature} $T_{\textit{ef{}f}}$, both of which are functions of the temperature and the chemical potential. Note that the parameter $\varepsilon$ is related to the expectation value of the component $T_{00}$ of the energy momentum tensor as described in \cite{Kundu:2016dyk}. To this end we first review the perturbative calculation of the area of the RT surfaces required for the computation of the entanglement negativity in subsection \ref{s2areahd} before proceeding to the non extremal bulk $AdS_{d+1}$-RN black holes in subsection \ref{s2hdne} and finally describing the corresponding bulk extremal $AdS_{d+1}$-RN black holes in subsection \ref{s2hdex}.

\subsection{Area of RT surfaces in $AdS_{d+1}$-RN}
\label{s2areahd}

The metric for the $AdS_{d+1}$-RN black hole (with $d\geq 3$) may be expressed as
\begin{equation}
\label{emetrichd}
ds^2=\frac{1}{z^2}\left(-f(z)dt^2+\frac{dz^2}{f(z)}+d\vec{x}^2\right),
\end{equation}
\\
\begin{equation}
f(z)=1-Mz^d+\frac{(d-2)Q^2}{(d-1)}z^{2(d-1)}\;,\qquad\qquad A_t(z)=Q\left(z_H^{d-2}-z^{d-2}\right),\nonumber
\end{equation}
where the $AdS$ length scale has been set to unity, and $Q$ and $M$ are the mass and charge of the black hole respectively, and $f$ is the lapse function. The horizon $z_H$ is determined by the smallest real root of the equation $f(z)=0$. The chemical potential $\mu $, conjugate to the charge $Q$, is then defined as follows
\begin{equation}
\label{ecpdef}
\mu \equiv\lim_{z\to 0}A_t(z)=Qz_H^{d-2},
\end{equation}
while the Hawking temperature is
\begin{equation}
\label{etemphd}
T=-\left.\frac{1}{4\pi}\frac{d}{dz}f(z)\right\rvert_{z=z_H}=\frac{d}{4\pi z_H}\left(1-\frac{(d-2)^2Q^2z_H^{2(d-1)}}{d(d-1)}\right).
\end{equation}
The total energy $\varepsilon$ of the system, which is a dimensionless quantity with limits $\frac{2(d-1)}{d-2}\leq \varepsilon\leq 1$, is given as follows \cite{Kundu:2016dyk}
\begin{equation}
\label{eephd}
\varepsilon(T,\mu )=b_0-\frac{2n}{1+\sqrt{1+\frac{d^2}{2\pi^2b_0b_1}\left(\frac{\mu^2}{T^2}\right)}}\;,
\end{equation}
where the numerical constants $b_0$ and $b_1$ are listed in appendix \ref{arnhd} in eqs.\@ \eqref{eb0} and \eqref{eb1} respectively.

The lapse function, chemical potential and temperature may then be expressed in terms of $\varepsilon$ as follows
\begin{equation}
f(z)=1-\varepsilon\left(\frac{z}{z_H}\right)^d+(\varepsilon-1)\left(\frac{z}{z_H}\right)^{2(d-1)},\nonumber
\end{equation}
\\
\begin{equation}
\mu =\frac{1}{z_H}\sqrt{\frac{(d-1)}{(d-2)}(\varepsilon-1)}\;\;,\nonumber
\end{equation}
\\
\begin{equation}
T=\frac{2(d-1)-(d-2)\varepsilon}{4\pi z_H}\;.
\end{equation}
The ef{}fective temperature $T_{\textit{ef{}f}}$, which describes the number of microstates for a given temperature and chemical potential, may be defined as \cite{Kundu:2016dyk}
\begin{equation}
\label{eteffhd}
T_{\textit{ef{}f}}(T,\mu )\equiv\frac{d}{4\pi z_H}=\frac{T}{2}\left[1+\sqrt{1+\frac{d^2}{2\pi^2b_0b_1}\left(\frac{\mu^2}{T^2}\right)}\;\right].
\end{equation}

We now proceed to review the computation the area of a RT surface homologous to a single subsystem $A$ with long rectangular strip geometry in a $CFT_d$ dual to a bulk $AdS_{d+1}$-RN black hole. The subsystem $A$ is specified by
\begin{equation}
\label{ecoordhda}
x\equiv x^1\in \left[-\frac{l}{2},\;\frac{l}{2}\right],\qquad x^i\in \left[-\frac{L}{2},\;\frac{L}{2}\right],\qquad i=2,\dots,d-2\,;
\end{equation}
with $l\ll L$. The area $\mathcal{A}_A$ of the RT surface homologous to this subsystem may then be expressed as
\begin{equation}
\label{eareahdxz}
\mathcal{A}_A=2L^{d-2}z_*^{d-1}\int_0^{l/2}\frac{dx}{z(x)^{2(d-1)}}=2L^{d-2}z_*^{d-1}\int_a^{z_*}\frac{dz}{z^{d-1}\sqrt{f(z)\left[z_*^{2(d-1)}-z^{2(d-1)}\right]}}\,,
\end{equation}
where $z_*$ is the turning point of the RT surface in the $x$ direction, and $a$ is the UV cut of{}f of the $CFT_d$. The length $l$ of the rectangular strip in the $x$ direction is related to $z_*$ as
\begin{equation}
\label{elenhdz}
l=2\int_0^{z_*}\frac{dz}{\sqrt{f(z)\left[\left(\frac{z_*}{z}\right)^{2(d-1)}-1\right]}}\,.
\end{equation}
The integral in eq.\@ \eqref{elenhdz} may be expressed as the following double sum \cite{Kundu:2016dyk}
\begin{equation}
\label{elenhddsum}
l=\frac{z_*}{d-1}\sum_{n=0}^\infty\sum_{k=0}^n\frac{\Gamma\left[\frac{1}{2}+n\right]\Gamma\left[\frac{d(n+k+1)-2k}{2(d-1)}\right]\varepsilon^{n-k}(1-\varepsilon)^k}{\Gamma[1+n-k]\Gamma[k+1]\Gamma\left[\frac{d(n+k+2)-2k-1}{2(d-1)}\right]}\left(\frac{z_*}{z_H}\right)^{nd+k(d-2)}.
\end{equation}
The area of the RT surface in eq.\@ \eqref{eareahdxz} may likewise be described as a double sum as follows \cite{Kundu:2016dyk}
\begin{equation}
\label{eareahddsum}
\begin{aligned}
\mathcal{A}_A&=\frac{2}{d-2}\left(\frac{L}{a}\right)^{d-2}+2\frac{L^{d-2}}{z_*^{d-2}}\left[\frac{\sqrt{\pi}\Gamma\left[-\frac{d-2}{2(d-1)}\right]}{2(d-1)\Gamma\left[\frac{1}{2(d-1)}\right]}\right]\\&+\frac{L^{d-2}}{(d-1)z_*^{d-2}}\left[\sum_{n=1}^\infty\sum_{k=0}^n\frac{\Gamma\left[\frac{1}{2}+n\right]\Gamma\left[\frac{d(n+k-1)-2k+2}{2(d-1)}\right]\varepsilon^{n-k}(1-\varepsilon)^k}{\Gamma[1+n-k]\Gamma[k+1]\Gamma\left[\frac{d(n+k)-2k+1}{2(d-1)}\right]}\left(\frac{z_*}{z_H}\right)^{nd+k(d-2)}\right].
\end{aligned}
\end{equation}

In what follows the area and the turning point of the corresponding bulk RT surface are expressed as perturbative expansions in terms of the specified parameters $\varepsilon$ and $T_{\textit{ef{}f}}$, for various limits of the temperature $T$ and the chemical potential $\mu $ of the dual $CFT_d$. The results are then utilized to compute the holographic entanglement negativity for various mixed state configurations of the single subsystem in the $CFT_d$s dual to bulk $AdS_{d+1}$-RN non extremal and extremal black holes, from the conjecture as described in eq.\@ \eqref{ehensg}. 

Here it is required to consider two auxiliary subsystems $B_1$ and $B_2$ on either side of the subsystem $A$ along the direction of the $x^1$ axis (henceforth referred to as the partitioning direction), specified by
\begin{equation}
\label{ecoordhdb1}
x^1\in \left[-\left(b+\frac{l}{2}\right),\;-\frac{l}{2}\right],\qquad x^i\in \left[-\frac{L}{2},\;\frac{L}{2}\right];
\end{equation}
and
\begin{equation}
\label{ecoordhdb2}
x^1\in \left[\frac{l}{2},\;\left(b+\frac{l}{2}\right)\right],\qquad x^i\in \left[-\frac{L}{2},\;\frac{L}{2}\right];
\end{equation}
respectively, where $i=2,\dots,d-2$, with $l\ll b$ and $b\ll L$. Here $L$ denotes the length of the strip in the remaining $(d-2)$ spatial directions. Note that we have chosen the subsystems $B_1$ and $B_2$ to be symmetric along the partitioning direction. This leads to the equality of the areas $\mathcal{A}_{B_1}=\mathcal{A}_{B_2}$ and $\mathcal{A}_{A\cup B_1}=\mathcal{A}_{A\cup B_2}$. With this identification, the holographic entanglement negativity for the above configuration may be computed from eq.\@ \eqref{ehensgsym}.

\subsection{Non extremal $AdS_{d+1}$-RN black holes}
\label{s2hdne}

In this subsection we utilize the area results for RT surfaces desribed above to compute the holographic entanglement negativity for finite temperature mixed states of single subsystems with a long rectangular strip geometry in $CFT_d$s dual to non extremal $AdS_{d+1}$-RN black holes for various non trivial limits of the temperature $T$ and the chemical potential $\mu $ describing dif{}ferent regimes. This involves the perturbative expansion of the areas of the bulk RT surfaces for appropriate combinations of the subsystem and the auxiliary subsystems, as mentioned in subsection \ref{s2areahd}. We begin with the regime of small chemical potential and low temperature in subsection \ref{s3hdnescplt}, proceeding to the case of small chemical potential and low temperature in subsection \ref{s3hdnescpht} and the large chemical potential and low temperature limit in subsection \ref{s3hdnelcplt}.

\subsubsection{Small chemical potential - low temperature}
\label{s3hdnescplt}

The regime of small chemical potential and low temperature is described by the conditions $\mu l\ll 1$ and $Tl\ll 1$. Depending on whether $T\ll\mu $ or $T\gg\mu $ \cite{Kundu:2016dyk}, this gives rise to two dif{}ferent situations as described below.
\bigskip

\noindent{\bfseries (i) $Tl\ll\mu l\ll 1$}
\bigskip

\noindent
We first consider the case where $ T\ll\mu $, which, together with $\mu l\ll 1$ and $Tl\ll 1$, may be recast as $Tl\ll\mu l\ll 1$. In this limit, the parameters $\varepsilon(T,\mu )$ and $T_{\textit{ef{}f}}(T,\mu )$, given by eqs.\@ \eqref{eephd} and eq.\@ \eqref{eteffhd}, may be expanded about $\frac{T}{\mu }=0$ to the leading order as \cite{Kundu:2016dyk}
\begin{equation}
\label{eephdnescpltst}
\varepsilon\approx b_0-\frac{2n\pi\sqrt{2b_0b_1}}{d}\left(\frac{T}{\mu }\right),
\end{equation}
\\
\begin{equation}
\label{eteffhdnescpltst}
T_{\textit{ef{}f}}\approx\frac{1}{2}\left(\frac{\mu d}{\pi\sqrt{2b_0b_1}}+T\right).
\end{equation}
In this regime the turning point of the bulk RT surface remains far away from the black hole horizon ($z_*\ll z_H$). Thus eq.\@ \eqref{elenhddsum} may be utilized to expand $z_*$ in terms of $\left(\frac{l}{z_H}\right)^d$ as
\begin{equation}
\label{ezshdnescpltst}
\begin{aligned}
z_*=\frac{l~\Gamma\left[\frac{1}{2(d-1)}\right]}{2\sqrt{\pi}~\Gamma\left[\frac{d}{2(d-1)}\right]}\left[1-\frac{1}{2(d+1)}\frac{2^{\frac{1}{d-1}-d}\Gamma\left[1+\frac{1}{2(d-1)}\right]\Gamma\left[\frac{1}{2(d-1)}\right]^{d+1}}{\pi^{\frac{d+1}{2}}\Gamma\left[\frac{1}{2}+\frac{1}{d-1}\right]\Gamma\left[\frac{d}{2(d-1)}\right]^d}\varepsilon~\left(\frac{l}{z_H}\right)^d\right]\\+\mathcal{O}\left[\left(\frac{l}{z_H}\right)^{2(d-1)}\right].
\end{aligned}
\end{equation}
The area of the RT surface homologous to a generic subsystem $\Sigma$ may likewise be computed from eq.\@ \eqref{eareahddsum} and re-expressed in terms of the parameters $\varepsilon$ and $T_{\textit{ef{}f}}$ as follows \cite{Kundu:2016dyk}
\begin{equation}
\label{eareahdnescpltst}
\mathcal{A}_{\Sigma}=\frac{2}{d-2}\left(\frac{L}{a}\right)^{d-2}+\mathcal{S}_0\left(\frac{L}{l}\right)^{d-2}+\varepsilon\mathcal{S}_0\mathcal{S}_1\left(\frac{4\pi T_{\textit{ef{}f}}}{d}\right)^dL^{d-2}l^2+\mathcal{O}\left[(T_{\textit{ef{}f}}l)^{2(d-1)}\right].
\end{equation}
where $l$ is the length of the subsystem $\Sigma$ in the partitioning direction, and the numerical constants $\mathcal{S}_0$ and $\mathcal{S}_1$ are listed in appendix \ref{arnhd} in eqs.\@ \eqref{eS0} and \eqref{eS1} respectively.

As discussed in \cite{Chaturvedi:2016rft}, the subsystems $B_1$ and $A\cup B_1$ in the $CFT_d$ become infinite along the direction of partition in the bipartite limit $B\to A^c$. Thus the RT surfaces homologous to these subsystems penetrate deep into the bulk to approach the black hole horizon ($z_*\approx z_H$). For these surfaces the appropriate Taylor expansion gives rise to the following expression for the area \cite{Kundu:2016dyk}
\begin{equation}
\label{eareaauxhdnescpltst}
\begin{aligned}
\mathcal{A}_{\Sigma}=\frac{2}{d-2}\left(\frac{L}{a}\right)^{d-2}&+\left(\frac{4\pi T_{\textit{ef{}f}}}{d}\right)^{d-1}L^{d-2}l\\&+\left(\frac{4\pi T_{\textit{ef{}f}}}{d}\right)^{d-2}\left.L^{d-2}\middle(N_0+N_1(b_0-\varepsilon)\right)+\mathcal{O}\left[\frac{T}{\mu }\right].
\end{aligned}
\end{equation}
The numerical constants $N_0$ and $N_1$ in eq.\@ \eqref{eareaauxhdnescpltst} are listed in appendix \ref{arnhd} in eqs.\@ \eqref{eN0} and \eqref{eN1} respectively.

Utilizing eqs.\@ \eqref{eareahdnescpltst} and \eqref{eareaauxhdnescpltst}, the holographic entanglement negativity in the limit of small chemical potential and low temperature (with $T\ll\mu $), for the mixed state configuration of the subsystem $A$ may now be obtained from the construction in \cite{Chaturvedi:2016rcn} by employing eq.\@ \eqref{ehensgsym} as follows\footnote{\label{nexphdnescpltst}Eq.\@ \eqref{eareahdnescpltst} is employed to calculate $\mathcal{A}_A$, while $\mathcal{A}_{B_1}$ and $\mathcal{A}_{A\cup B_1}$ are obtained utilizing eq.\@ \eqref{eareaauxhdnescpltst}, followed by implementation of the bipartite limit $l/b\to 0$.}
\begin{equation}
\label{ehenhdnescpltst}
\begin{aligned}
\mathcal{E}_A=\frac{3}{8G_N^{(d+1)}}\left[\frac{2}{d-2}\left(\frac{L}{a}\right)^{d-2}\right.&+\mathcal{S}_0\left(\frac{L}{l}\right)^{d-2}+\varepsilon\mathcal{S}_0\mathcal{S}_1\left(\frac{4\pi T_{\textit{ef{}f}}}{d}\right)^dL^{d-2}l^2\\&\left.-\left(\frac{4\pi T_{\textit{ef{}f}}}{d}\right)^{d-1}L^{d-2}l\right]+\mathcal{O}\left[(T_{\textit{ef{}f}}l)^{2(d-1)}\right].
\end{aligned}
\end{equation}
The corresponding holographic entanglement entropy is given from eq.\@ \eqref{eheesg} through the use of eq.\@ \eqref{eareahdnescpltst} as follows
\begin{equation}
\label{eheehdnescpltst}
\begin{aligned}
\mathcal{S}_A=\frac{1}{4G_N^{(d+1)}}\left[\frac{2}{d-2}\left(\frac{L}{a}\right)^{d-2}+\mathcal{S}_0\left(\frac{L}{l}\right)^{d-2}+\varepsilon\mathcal{S}_0\mathcal{S}_1\left(\frac{4\pi T_{\textit{ef{}f}}}{d}\right)^dL^{d-2}l^2\right]\\+\mathcal{O}\left[(T_{\textit{ef{}f}}l)^{2(d-1)}\right].
\end{aligned}
\end{equation}
Comparing eq.\@ \eqref{ehenhdnescpltst} with eq.\@ \eqref{eheehdnescpltst} we arrive at
\begin{equation}
\label{ehenheehdnescpltst}
\mathcal{E}_A=\left.\frac{3}{2}\middle(\mathcal{S}_A-\mathcal{S}_A^{\textit{th}}\right),
\end{equation}
where $\mathcal{S}_A^{\textit{th}}=\frac{1}{4G_N^{(d+1)}}\left(\frac{4\pi T_{\textit{ef{}f}}}{d}\right)^{d-1}V$ with $V\equiv L^{d-2}l$ being the volume of the subsystem under consideration.

The first two terms on the right hand side of eq.\@ \eqref{ehenhdnescpltst} for the holographic entanglement negativity correspond to the holographic entanglement negativity for the zero temperature pure state configuration of a single subsystem in a $CFT_d$ dual to bulk pure $AdS_{d+1}$ geometry, while the rest of the terms characterize the perturbative corrections due to the temperature and chemical potential of the black hole.
The relation in eq.\@ \eqref{ehenheehdnescpltst} describes that the holographic entanglement negativity is characterized at the leading order by the elimination of the volume dependent thermal entropy from the holographic entanglement entropy for the finite temperature mixed state of a single subsystem reproducing earlier results in the literature. Note that this is in contrast with the holographic entanglement entropy given in eq.\@ \eqref{eheehdnescpltst}, which scales with the volume. This is expected as the entanglement negativity characterizes an upper bound on the distillable entanglement in quantum information theory \cite{Vidal:2002zz,Plenio:2005cwa}. The elimination of the thermal contributions to the leading order hence serves as a strong consistency check for the higher dimensional holographic construction described in \cite{Chaturvedi:2016rcn, Chaturvedi:2016rft}.\footnote{This elimination of the thermal contributions is expected to hold to all orders. Note also that an exact analysis of the area of the RT surfaces is there in the literature as mentioned in footnote \ref{nmeijer}.} Note that this is similar to the corresponding lower dimensional $AdS_3/CFT_2$ scenario where it is exact in the large central charge limit.

\bigskip

\noindent{\bfseries (ii) $\mu l\ll Tl\ll 1$}
\bigskip

\noindent
Next we consider the limit $T\gg\mu $. This condition may be combined with $\mu l\ll 1$ and $Tl\ll 1$ into $\mu l\ll Tl\ll 1$. For this case the parameters $\varepsilon(T,\mu )$ and $T_{\textit{ef{}f}}(T,\mu )$ described by eqs.\@ \eqref{eephd} and \eqref{eteffhd} have the following Taylor expansions around $\frac{\mu }{T}=0$ \cite{Kundu:2016dyk}
\begin{equation}
\label{eephdnescpltbt}
\varepsilon=1+\frac{d^2(d-2)}{16\pi^2(d-1)}\left(\frac{\mu }{T}\right)^2+\mathcal{O}\left[\left(\frac{\mu }{T}\right)^4\right],
\end{equation}
\begin{equation}
\label{eteffhdnescpltbt}
T_{\textit{ef{}f}}=T\left[1+\frac{d(d-2)^2}{16\pi^2(d-1)}\left(\frac{\mu }{T}\right)^2+\mathcal{O}\left[\left(\frac{\mu }{T}\right)^4\right]\right].
\end{equation}
Once again the conditions $Tl\ll 1$ and $\mu l\ll 1$ entail that $z_*\ll z_H$. The expression for $z_*$ may be obtained by expanding eq.\@ \eqref{elenhddsum} in terms of $\left(\frac{l}{z_H}\right)^d$, and is same as the one given in eq.\@ \eqref{ezshdnescpltst}. The area of the RT surface homologous to a generic subsystem $\Sigma$ is once again determined by expansion of eq.\@ \eqref{eareahddsum} in terms of $\left(\frac{l}{z_H}\right)^d $ as \cite{Kundu:2016dyk}
\begin{equation}
\label{eareahdnescpltbt}
\mathcal{A}_{\Sigma}=\frac{2}{d-2}\left(\frac{L}{a}\right)^{d-2}+\mathcal{S}_0\left(\frac{L}{l}\right)^{d-2}+\varepsilon\mathcal{S}_0\mathcal{S}_1\left(\frac{4\pi T_{\textit{ef{}f}}}{d}\right)^dL^{d-2}l^2+\mathcal{O}\left[(T_{\textit{ef{}f}}l)^{2(d-1)}\right],
\end{equation}
where the numerical constants $\mathcal{S}_0$ and $\mathcal{S}_1$ are the same as in eq.\@ \eqref{eareahdnescpltst}, and are given in appendix \ref{arnhd} in eqs.\@ \eqref{eS0} and \eqref{eS1} respectively.

We note here that although the areas of the bulk RT surface given in eqs.\@ \eqref{eareahdnescpltst} and \eqref{eareahdnescpltbt} have identical forms for both $Tl\ll\mu l\ll 1$ and $\mu l\ll Tl\ll 1$, the underlying parameters $\varepsilon$ and $T_{\textit{ef{}f}}$ are dif{}ferent for the two cases as described in eqs.\@ \eqref{eephdnescpltst} and \eqref{eteffhdnescpltst}, and eqs.\@ \eqref{eephdnescpltbt} and \eqref{eteffhdnescpltbt} respectively. As outlined in the discussion just preceding eq.\@ \eqref{eareaauxhdnescpltst}, the subsystems $B_1$ and $A\cup B_1$ become infinite along the partitioning direction in the limit $B\to A^c$, and consequently the corresponding RT surfaces extend deep into the bulk so that the turning points of these surfaces are close to the horizon ($z_*\approx z_H$). The relevant expression for the area of these surfaces is given as follows \cite{Kundu:2016dyk}
\begin{equation}
\label{eareaauxhdnescpltbt}
\mathcal{A}_\Sigma=\frac{2}{d-2}\left(\frac{L}{a}\right)^{d-2}+\left(\frac{4\pi T_{\textit{ef{}f}}}{d}\right)^{d-1}L^{d-2}l+\left(\frac{4\pi T_{\textit{ef{}f}}}{d}\right)^{d-2}L^{d-2}\gamma_d\left(\frac{\mu }{T}\right).
\end{equation}
The function $\gamma_d\left(\frac{\mu }{T}\right)$ in eq.\@ \eqref{eareaauxhdnescpltbt} admits a perturbative expansion in $\frac{\mu }{T}$, which has been described in appendix \ref{arnhd} in eq.\@ \eqref{gamma}.

The holographic entanglement negativity for the required finite temperature mixed state configuration of the single subsystem in question with long rectangular strip geometry, in the limit of small chemical potential and low temperature (with $T\gg\mu $) may then be computed\footnote{Once again $\mathcal{A}_A$ is computed by applying eq.\@ \eqref{eareahdnescpltbt}, whereas eq.\@ \eqref{eareaauxhdnescpltbt} is utilized to obtain $\mathcal{A}_{B_1}$ and $\mathcal{A}_{A\cup B_1}$, and afterwards we take the bipartite limit $l/b\to 0$.} from eqs.\@ \eqref{eareahdnescpltbt} and \eqref{eareaauxhdnescpltbt}, from the conjecture described by eq.\@ \eqref{ehensgsym}, and has a form identical to that in eq.\@ \eqref{ehenhdnescpltst}. Relations described in eqs.\@ \eqref{eheehdnescpltst} and \eqref{ehenheehdnescpltst}, and the discussions after eq.\@ \eqref{ehenheehdnescpltst} also follow identically.

\subsubsection{Small chemical potential - high temperature}
\label{s3hdnescpht}

In this subsection we proceed to investigate the holographic entanglement negativity in the limit of small chemical potential and high temperature. As detailed in \cite{Kundu:2016dyk}, this regime is given by the conditions $\mu \ll T$ and $Tl\gg 1$. The parameters $\varepsilon$ and $T_{\textit{ef{}f}}$ may be given in the limit $\mu \ll T$ (refer to case (ii), subsection \ref{s3hdnescplt} for details) by eqs.\@ \eqref{eephdnescpltbt} and \eqref{eteffhdnescpltbt}. The other condition $Tl\gg 1$ implies $z_*\approx z_H$. Thus the area of the bulk extremal surface may be obtained by perturbative expansion of eq.\@ \eqref{eareahddsum} around $\frac{z_*}{z_H}=1$, and is given by eq.\@ \eqref{eareaauxhdnescpltbt}.

Substituting eq.\@ \eqref{eareaauxhdnescpltbt} into eq.\@ \eqref{ehensgsym}, the holographic entanglement negativity in the limit of small chemical potential and high temperature, for the mixed state configuration of the single subsystem under consideration may be determined as follows
\begin{equation}
\label{ehenhdnescpht}
\mathcal{E}_A=\frac{3}{8G_N^{(d+1)}}\left[\frac{2}{d-2}\left(\frac{L}{a}\right)^{d-2}+\left(\frac{4\pi T_{\textit{ef{}f}}}{d}\right)^{d-2}L^{d-2}\gamma_d\left(\frac{\mu }{T}\right)\right].
\end{equation}
Next we employ eq.\@ \eqref{eareaauxhdnescpltbt} to compute the corresponding holographic entanglement entropy from eq.\@ \eqref{eheesg} as follows
\begin{equation}
\label{eheehdnescpht}
\mathcal{S}_A=\frac{1}{4G_N^{(d+1)}}\left[\frac{2}{d-2}\left(\frac{L}{a}\right)^{d-2}+\left(\frac{4\pi T_{\textit{ef{}f}}}{d}\right)^{d-1}L^{d-2}l+\left(\frac{4\pi T_{\textit{ef{}f}}}{d}\right)^{d-2}L^{d-2}\gamma_d\left(\frac{\mu }{T}\right)\right].
\end{equation}
Comparing eqs.\@ \eqref{ehenhdnescpht} and \eqref{eheehdnescpht} we again obtain
\begin{equation}
\label{ehenheehdnescpht}
\mathcal{E}_A=\left.\frac{3}{2}\middle(\mathcal{S}_A-\mathcal{S}_A^{\textit{th}}\right),
\end{equation}
where $\mathcal{S}_A^{\textit{th}}=\frac{1}{4G_N^{(d+1)}}\left(\frac{4\pi T_{\textit{ef{}f}}}{d}\right)^{d-1}L^{d-2}l$ as before.

Once again we observe that the expression for the holographic entanglement negativity to the leading order in eq.\@ \eqref{ehenhdnescpht} for the mixed state in question is purely dependent on the area $L^{d-2}$ of the entangling surface between the single subsystem with the rest of the system. As earlier this indicates the vanishing of the volume dependent thermal contribution, as may be seen from eq.\@ \eqref{ehenheehdnescpht}.

\subsubsection{Large chemical potential - low temperature}
\label{s3hdnelcplt}

The case for large chemical potential and low temperature regime is described by the conditions $\mu l\gg 1$ and $T\ll\mu $. As mentioned earlier (case (i), subsection \ref{s3hdnescplt}), the condition $T\ll\mu $ implies that the parameters $\varepsilon(T,\mu )$ and $T_{\textit{ef{}f}}(T,\mu )$ admit of the expansions described by eqs.\@ \eqref{eephdnescpltst} and \eqref{eteffhdnescpltst} respectively. Similar to subsection \ref{s3hdnescpht}, the condition $\mu l\gg 1$ dictates $z_*\approx z_H$. Hence, the area of the bulk RT surface may once again be obtained through the perturbative expansion of eq.\@ \eqref{eareahddsum} around $\frac{z_*}{z_H}=1$ \cite{Kundu:2016dyk}, and is given by eq.\@ \eqref{eareaauxhdnescpltst}.

Utilizing eq.\@ \eqref{eareaauxhdnescpltst}, we may now obtain the holographic entanglement negativity in the limit of large chemical potential and low temperature, for the mixed state configuration of the subsystem $A$, from the construction advanced in \cite{Chaturvedi:2016rcn} by employing eq.\@ \eqref{ehensgsym} as follows
\begin{equation}
\label{ehenhdnelcplt}
\mathcal{E}_A=\frac{3}{8G_N^{(d+1)}}\left[\frac{2}{d-2}\left(\frac{L}{a}\right)^{d-2}+\left(\frac{4\pi T_{\textit{ef{}f}}}{d}\right)^{d-2}\left.L^{d-2}\middle(N_0+N_1(b_0-\varepsilon)\right)\right]+\mathcal{O}\left[\frac{T}{\mu }\right].
\end{equation}
Similarly the entanglement entropy for this regime may also be determined from eq.\@ \eqref{eheesg} using eq.\@ \eqref{eareaauxhdnescpltst} as follows
\begin{equation}
\label{eheehdnelcplt}
\begin{aligned}
\mathcal{S}_A=\frac{1}{4G_N^{(d+1)}}\left[\frac{2}{d-2}\left(\frac{L}{a}\right)^{d-2}\right.&+\left(\frac{4\pi T_{\textit{ef{}f}}}{d}\right)^{d-1}L^{d-2}l\\&\left.+\left(\frac{4\pi T_{\textit{ef{}f}}}{d}\right)^{d-2}\left.L^{d-2}\middle(N_0+N_1\left(b_0-\varepsilon\right)\right)\right]+\mathcal{O}\left[\frac{T}{\mu }\right].
\end{aligned}
\end{equation}
Once again comparison of eqs.\@ \eqref{ehenhdnelcplt} and \eqref{eheehdnelcplt} leads us to
\begin{equation}
\label{ehenheehdnelcplt}
\mathcal{E}_A=\left.\frac{3}{2}\middle(\mathcal{S}_A-\mathcal{S}_A^{\textit{th}}\right),
\end{equation}
where as earlier $\mathcal{S}_A^{\textit{th}}=\frac{1}{4G_N^{(d+1)}}\left(\frac{4\pi T_{\textit{ef{}f}}}{d}\right)^{d-1}L^{d-2}l$.

Again the holographic entanglement negativity to the leading order in eq.\@ \eqref{ehenhdnelcplt}, obtained in the limit of large chemical potential and high temperature, is purely dependent on the area of the relevant entangling surface. As earlier this is due to the precise elimination of the volume dependent thermal term to the leading order described by eq.\@ \eqref{ehenheehdnelcplt}, in conformity with the standard quantum information theory expectations, and serves to further substantiate the holographic conjecture proposed in \cite{Chaturvedi:2016rcn} and our construction.

\subsection{Extremal $AdS_{d+1}$-RN black holes}
\label{s2hdex}

In this subsection we outline the procedure to obtain the holographic entanglement negativity for the zero temperature mixed state configurations of single subsystems in $CFT_d$s dual to bulk extremal $AdS_{d+1}$-RN black holes. The relevant parameters for these bulk extremal black hole geometries are given as \cite{Kundu:2016dyk}
\begin{equation}
\label{eqhdex}
Q^2=\frac{d(d-1)L^2}{(d-2)^2z_H^{2(d-1)}}\;,
\end{equation}
\begin{equation}
\varepsilon=b_1,
\end{equation}
\begin{equation}
\label{ecphdex}
\mu =\frac{1}{z_H}\sqrt{\frac{b_0b_1}{2}}=\frac{1}{z_H}\sqrt{\frac{d(d-1)}{(d-2)^2}}\;,
\end{equation}
\begin{equation}
T_{\textit{ef{}f}}=\frac{\mu d}{2\pi\sqrt{2b_0b_1}}\;.
\end{equation}
In the above equations the parameter $Q$ describes the charge of the extremal $AdS_{d+1}$-RN black hole, while $T_{\textit{ef{}f}}$ as earlier denotes the ef{}fective temperature.

In the next two subsections we proceed to obtain perturbative expansions of the area of the RT surface homologous to the single subsystem with long rectangular strip geometry, for two dif{}ferent limits of the charge $Q$. The expression obtained will then be utilized to compute the holographic entanglement negativity for the mixed state configuration under consideration from the conjecture proposed in \cite{Chaturvedi:2016rcn}. In subsection \ref{s3hdexscp} we study the case involving a small chemical potential, while subsection \ref{s3hdexlcp} describes the large chemical potential regime.

\subsubsection{Small chemical potential}
\label{s3hdexscp}

The limit of small chemical potential is defined by the condition $\mu l\ll 1$, which entails $z_*\ll z_H$. The solution of eq.\@ \eqref{elenhdz} in this limit once again leads to eq.\@ \eqref{ezshdnescpltst} \cite{Kundu:2016dyk}. The area of the RT surface homologous to a generic subsystem $\Sigma$ may be similarly obtained from eq.\@ \eqref{eareahddsum} and re-expressed in terms $\mu $ as follows \cite{Kundu:2016dyk}
\begin{equation}
\label{eareahdexscp}
\begin{aligned}
\mathcal{A}_\Sigma=\frac{2}{d-2}\left(\frac{L}{a}\right)^{d-2}+\mathcal{S}_0\left(\frac{L}{l}\right)^{d-2}+\frac{2(d-1)}{d-2}\mathcal{S}_0\mathcal{S}_1\left(\frac{\mu (d-2)}{\sqrt{d(d-1)}}\right)^dL^{d-2}l^2\\+\mathcal{O}\left[(\mu l)^{2(d-1)}\right],
\end{aligned}
\end{equation}
where the constants $\mathcal{S}_0$ and $\mathcal{S}_1$ are identical to those in the earlier cases described in subsection \ref{s3hdnescplt}, and are given in appendix \ref{arnhd} in eqs.\@ \eqref{eS0} and \eqref{eS1} respectively.

As discussed in earlier subsections, the turning points of the RT surfaces corresponding to the infinite auxiliary subsystems approach the black hole horizon ($z_*\approx z_H$). The appropriate area expansion for these surfaces are given as \cite{Kundu:2016dyk}
\begin{equation}
\label{eareaauxhdexscp}
\mathcal{A}_\Sigma=\frac{2}{d-2}\left(\frac{L}{a}\right)^{d-2}+\left(\frac{\mu (d-2)}{\sqrt{d(d-1)}}\right)^{d-1}L^{d-2}l+\left(\frac{\mu (d-2)}{\sqrt{d(d-1)}}\right)^{d-2}L^{d-2}N(b_0),
\end{equation}
where $N(b_0)$ is the value of $N(\varepsilon)$ at $\varepsilon=b_0$. It is listed in appendix \ref{arnhd} in eq.\@ \eqref{eNep}.

Utilizing eqs.\@ \eqref{eareahdexscp} and \eqref{eareaauxhdexscp}, it is now possible to compute the holographic entanglement negativity in the limit of small chemical potential, for the bipartite configuration of the single subsystem in question by employing eq.\@ \eqref{ehensgsym} as follows
\begin{equation}
\label{ehenhdexscp}
\begin{aligned}
\mathcal{E}_A=\frac{3}{8G_N^{(d+1)}}\left[\frac{2}{d-2}\left(\frac{L}{a}\right)^{d-2}+\mathcal{S}_0\left(\frac{L}{l}\right)^{d-2}+\frac{2(d-1)}{d-2}\mathcal{S}_0\mathcal{S}_1\left(\frac{\mu (d-2)}{\sqrt{d(d-1)}}\right)^dL^{d-2}l^2\right.\\\left.-\left(\frac{\mu (d-2)}{\sqrt{d(d-1)}}\right)^{d-1}L^{d-2}l\right]+\mathcal{O}\left[(\mu l)^{2(d-1)}\right].
\end{aligned}
\end{equation}
We then employ eq.\@ \eqref{eareahdexscp} to obtain the following expression for the holographic entanglement entropy from eq.\@ \eqref{eheesg}
\begin{equation}
\label{eheehdexscp}
\begin{aligned}
\mathcal{S}_A=\frac{1}{4G_N^{(d+1)}}\left[\frac{2}{d-2}\left(\frac{L}{a}\right)^{d-2}+\mathcal{S}_0\left(\frac{L}{l}\right)^{d-2}+\frac{2(d-1)}{d-2}\mathcal{S}_0\mathcal{S}_1\left(\frac{\mu (d-2)}{\sqrt{d(d-1)}}\right)^dL^{d-2}l^2\right]\\+\mathcal{O}\left[(\mu l)^{2(d-1)}\right].
\end{aligned}
\end{equation}
Through the comparison of eqs.\@ \eqref{ehenhdexscp} and \eqref{eheehdexscp} we arrive at the following relation
\begin{equation}
\label{ehenheehdexscp}
\mathcal{E}_A=\left.\frac{3}{2}\middle(\mathcal{S}_A-\mathcal{S}_A^{\textit{th}}\right),
\end{equation}
where we have $\mathcal{S}_A^{\textit{th}}=\frac{1}{4G_N^{(d+1)}}\left(\frac{\mu (d-2)}{\sqrt{d(d-1)}}\right)^{d-1}L^{d-2}l$.

In eq.\@ \eqref{ehenhdexscp} we note that the first two terms describe the holographic entanglement negativity for the zero temperature pure state configuration of a single subsystem with long rectangular strip geometry in a $CFT_d$ dual to bulk pure $AdS_{d+1}$ geometry. The rest of the terms represent the corrections due to the chemical potential of the black hole. Once again we observe the elimination of the volume dependent {\it ``thermal''} contribution to the leading order\footnote{The thermal contribution characterizes the entropy due to the ground state degeneracy of the black hole, corresponding to the vacuum degeneracy of the dual $CFT$.} in eq.\@ \eqref{ehenheehdexscp} which is in conformity with quantum information theory and substantiates the holographic conjecture in \cite{Chaturvedi:2016rcn}.

\subsubsection{Large chemical potential}
\label{s3hdexlcp}

In this subsection we investigate the large chemical potential regime, specified by the condition $\mu l\gg 1$. With this condition it may be observed from eq.\@ \eqref{ecphdex} that the horizon radius is large, and the turning point of the bulk RT surface is thus close to the horizon ($z_*\approx z_H$). In this regime, the area of the RT surface homologous to a generic subsystem $\Sigma$ with long rectangular strip geometry may be determined by a perturbative computation of the integral in eq.\@ \eqref{eareahddsum} around $\frac{z_*}{z_H}=1$, and is given by eq.\@ \eqref{eareaauxhdexscp}.

The holographic entanglement negativity in the limit of large chemical potential, for the bipartite configuration of the single subsystem under consideration may now be derived through the use of eq.\@ \eqref{eareaauxhdexscp} by employing eq.\@ \eqref{ehensgsym} as follows
\begin{equation}
\label{ehenhdexlcp}
\mathcal{E}_A=\frac{3}{8G_N^{(d+1)}}\left[\frac{2}{d-2}\left(\frac{L}{a}\right)^{d-2}+\left(\frac{\mu (d-2)}{\sqrt{d(d-1)}}\right)^{d-2}L^{d-2}N(b_0)\right].
\end{equation}
The corresponding expression for the holographic entanglement entropy may then be obtained from eq.\@ \eqref{eareaauxhdexscp} utilizing eq.\@ \eqref{eheesg} as
\begin{equation}
\label{eheehdexlcp}
\mathcal{S}_A=\frac{1}{4G_N^{(d+1)}}\left[\frac{2}{d-2}\left(\frac{L}{a}\right)^{d-2}+\left(\frac{\mu (d-2)}{\sqrt{d(d-1)}}\right)^{d-1}L^{d-2}l+\left(\frac{\mu (d-2)}{\sqrt{d(d-1)}}\right)^{d-2}L^{d-2}N(b_0)\right].
\end{equation}
Once again comparing eqs.\@ \eqref{ehenhdexlcp} and \eqref{eheehdexlcp} we arrive at
\begin{equation}
\label{ehenheehdexlcp}
\mathcal{E}_A=\left.\frac{3}{2}\middle(\mathcal{S}_A-\mathcal{S}_A^{\textit{th}}\right),
\end{equation}
where as before we have identified $\mathcal{S}_A^{\textit{th}}=\frac{1}{4G_N^{(d+1)}}\left(\frac{\mu (d-2)}{\sqrt{d(d-1)}}\right)^{d-1}L^{d-2}l$ as the {\it ``thermal''} entropy.

We once again note that the holographic entanglement negativity computed in eq.\@ \eqref{ehenhdexlcp} for the large chemical potential regime is purely dependent on the area of the appropriate entangling surface, due to the elimination of the volume dependent {\it ``thermal''} contribution to the leading order described in eq.\@ \eqref{ehenheehdexlcp}, which agrees with quantum information theory expectations and substantiates the holographic construction in \cite{Chaturvedi:2016rcn} as earlier.

\subsection{Discussion of the results}
\label{s2disc}

The leading order results derived in sections \ref{s2hdne} and \ref{s2hdex} for the holographic entanglement negativity of a single subsystem in $CFT_d$s dual to bulk non extremal and extremal $AdS_{d+1}$-RN black holes for various limits of the relevant parameters are consistent with the quantum information theory expectations. It is observed that in the limit of small chemical potential and low temperature the leading contribution to the holographic entanglement negativity arises from the zero temperature configuration of single subsystem in the $CFT_d$ dual to the bulk pure $AdS_{d+1}$ geometry, while the correction terms depend on the chemical potential $\mu$ and the temperature $T$. The reason for this is that the bulk RT surface is located near the boundary far away from the black hole horizon, and thus the leading contribution arises from the near boundary pure $AdS_{d+1}$ geometry.

On the other hand for both the regimes of small chemical potential and high temperature, and large chemical potential and low temperature, the leading part is purely dependent on the area of the entangling surface between the subsystem and the rest of the system, while the volume dependent thermal term is eliminated. In this case the dominant contribution arises from the entanglement between the degrees of freedom at the entangling surface. For the case of small chemical potential with extremal $AdS_{d+1}$-RN black hole, the leading contribution involves the bulk pure $AdS_{d+1}$ geometry term, and correction terms dependent on the chemical potential $\mu$. However for the large charge regime, the leading contribution arises only from the terms dependent purely on the area of the relevant entangling surface, while once again the volume dependent thermal term drops out.

From the above description we note that the holographic entanglement negativity to the leading order is purely a function of the area of the entangling surface in the $CFT$ even in the regimes of large charge or high temperature, whereas the corresponding holographic entanglement entropy involves volume dependent thermal contributions. This ensues from the precise cancellation of the volume dependent thermal terms in the expressions for the holographic entanglement negativity to the leading order and is expected to hold to all orders. As discussed in the previous sections, this elimination reproduces the corresponding phenomena observed also for the $AdS_3/CFT_2$ scenario where this elimination is exact in the large central charge limit and is expected to hold to all orders in the perturbation theory. Naturally these results provide strong consistency checks, and serve to further substantiate the higher dimensional holographic construction proposed in \cite{Chaturvedi:2016rcn}.

\section{Summary and conclusions}
\label{ssum}

To summarize, we have extended the higher dimensional holographic entanglement negativity proposal for bipartite states described by a single subsystem in $CFT_d$s dual to bulk $AdS_{d+1}$ geometries described in \cite{Chaturvedi:2016rft} to $CFT_d$s with a conserved charge dual to bulk $AdS_{d+1}$-RN black holes. Our construction involved a specific algebraic sum of the areas of RT surfaces for specific combinations of the subsystem and auxiliary subsystems in the dual $CFT_d$.\footnote{\label{nsumblkprf}A plausible proof for the holographic construction from a gravitational path integral perspective in the literature has been reviewed in appendix \ref{ahendrv}.} In this context we have computed the holographic entanglement negativity for such bipartite states described by single subsystems with long rectangular strip geometries in $CFT_d$s with a conserved charge dual to bulk non extremal and extremal $AdS_{d+1}$-RN black holes.

To elucidate our construction we have reviewed the perturbative techniques utilized to compute the areas of the relevant bulk RT surfaces for combinations of subsystems in the dual $CFT_d$ with a conserved charge as mentioned above. To establish the perturbative techniques in this context, we have described various regimes characterized by the total energy and an ef{}fective temperature which are functions of the temperature and the chemical potential conjugate to the charge of the bulk RN black hole in the $AdS_{d+1}/CFT_d$ scenario. We have obtained the holographic entanglement negativity for bipartite states of single subsystems with long rectangular strip geometries for the regimes described above in $CFT_d$s with a conserved charge dual to bulk non extremal and extremal $AdS_{d+1}$-RN black holes.

Interestingly for all the cases we have demonstrated the elimination of the thermal contributions for the holographic entanglement negativity at the leading order which is similar to the case of $CFT_d$s dual to bulk $AdS_{d+1}$-Schwarzschild black holes described earlier in the literature. This feature for the holographic entanglement negativity is also observed for the corresponding $AdS_3/CFT_2$ scenario where it is exact in the large central charge limit. The elimination of the thermal contributions to the leading order mentioned above is expected to hold to all orders in the perturbation theory. This also conforms to quantum information theory expectations for the entanglement negativity which characterizes an upper bound on the distillable entanglement of a mixed state. Furthermore, this serves as a strong non trivial consistency check for the higher dimensional holographic construction for the entanglement negativity as proposed and exemplified in the literature.

We emphasize here that there is an alternative holographic entanglement negativity proposal involving the area of a backreacted minimal entanglement wedge cross section due to a cosmic brane for the conical defect of the bulk replicated geometry. For subsystems with spherical entangling surfaces this backreaction is described by an overall dimension dependent numerical constant which may be explicitly computed. The equivalence of this construction modulo certain constants, with the proposal involving the algebraic sums of the lengths of bulk geodesics for combinations of intervals has been explicitly demonstrated in the literature for the $AdS_3/CFT_2$ scenario.

Keeping the above discussion in perspective, the holographic construction and the results described in the present article are also expected to involve appropriate backreaction ef{}fects. However for the rectangular strip subsystem geometries the explicit computation of these ef{}fects remains a non trivial open issue. As mentioned in footnote \ref{nsumblkprf} a plausible bulk proof of our construction for spherical entangling surfaces in the literature has been reviewed in appendix \ref{ahendrv}. However the extension of this proof to general subsystem geometries is an outstanding issue.

Our results described in this article serve to further substantiate the holographic entanglement negativity proposal for bipartite states in $CFT_d$s in the context of a generic higher dimensional $AdS_{d+1}/CFT_d$ scenario which may find interesting applications to the study of diverse issues involving mixed state entanglement for various phenomena ranging from many body systems to quantum gravity and black holes. We emphasize that our construction involving the algebraic sums of the areas of RT surfaces may be simpler than determining the EWCS in higher dimensions which is extremely non trivial. Furthermore our results are also expected to provide insight into the generalization of the bulk proof from a gravitational path integral obtained for spherical entangling surfaces to subsystems with arbitrary geometries, which is an extremely dif{}ficult outstanding open problem. We hope to address these interesting issues in the near future.

\section*{Acknowledgments}

We would like to thank Vinay Malvimat and Debarshi Basu for crucial suggestions and discussions. The work of Sayid Mondal is supported in part by the Ministry of Science and Technology of Taiwan (under Grant No.\@ 106-2112-M-033-007-MY3 and Grant No.\@ 109-2112-M-033-005-MY3) and by the National Center for Theoretical Sciences (NCTS) of Taiwan.

\begin{appendices}

\section{Derivation of holographic entanglement negativity}
\label{ahendrv}

In this appendix we follow \cite{KumarBasak:2020ams} to provide a brief review of a plausible proof of the holographic conjecture for the entanglement negativity of a single interval in a $(1+1)$ dimensional conformal field theory ($CFT_{1+1}$) dual to a bulk $AdS_3$ geometry. As described in section \ref{shenr} this involves a specific algebraic sum of the lengths (areas) of bulk space like geodesics (RT surfaces). This proof follows from \cite{Dong:2021clv} which provided the areas of backreacting cosmic branes through the analysis of replica symmetry breaking saddles for the bulk gravitational path integral.

\subsection{Holographic R\'{e}nyi entanglement entropy}
\label{a2hree}

We first breifly review the holographic characterization of the R\'{e}nyi entropy following \cite{Dong:2016fnf} in this section. In this work the gravitational path integral technique in \cite{Lewkowycz:2013nqa} was employed to establish the relation between the holographic R\'{e}nyi entropy of order $n$ of a subsystem in a dual CFT and the area of the corresponding bulk cosmic brane $\mathcal{C}_n$ homologous to the subsystem as
\begin{equation}
\label{ehrecb}
n^2\frac{\partial}{\partial n}\left(\frac{n-1}{n}S^{(n)}(A)\right)=\frac{\mathrm{Area}\left(\mathcal{C}_n\right)}{4G_N}\,,
\end{equation}
where $G_N$ is the Newton constant in appropriate bulk dimensions. Next we define a quantity $\mathcal{A}^{(n)}$ in terms of the $n^{th}$ holographic R\'{e}nyi entropy $S^{(n)}(A)$ as follows
\begin{equation}
\label{ehrearea}
S^{(n)}(A)=\frac{\mathcal{A}^{(n)}}{4G_N}\,.
\end{equation}
Substituting eq.\@ \eqref{ehrearea} into eq.\@ \eqref{ehrecb}, we find the relation between $\mathcal{A}^{(n)}$ and the area of the cosmic brane as
\begin{equation}
\label{eareacb}
n^2\frac{\partial}{\partial n}\left(\frac{n-1}{n}\mathcal{A}^{(n)}\right)=\mathrm{Area}\left(\mathcal{C}_n\right).
\end{equation}
In the following analysis we will refer to $\mathcal{A}^{(n)}$ as the area of the backreacting cosmic brane $\mathcal{C}_n$ but it is to be understood in the context of eq.\@ \eqref{eareacb}. The tension $T_n$ of the cosmic brane $\mathcal{C}_n$ is related to the replica index $n$ as
\begin{equation}
\label{etencb}
T_n=\frac{n-1}{4nG_N}\,.
\end{equation}
Note that the tension $T_n$ of the cosmic brane $\mathcal{C}_n$ in eq.\@ \eqref{etencb} vanishes in the replica limit $n\to 1$, and the holographic construction in eq.\@ \eqref{ehrecb} reduces to the Ryu-Takayanagi (RT) conjecture. For $n\neq 1$ however the tension in eq.\@ \eqref{etencb} is non zero, which leads to a non trivial backreaction of the brane on the replicated bulk geometry. In this case the corresponding area of the cosmic brane $\mathcal{C}_n$ is dif{}ficult to compute. Remarkably the backreaction and the correspondng area of the cosmic brane were explicitly computed for subsystems with spherical entangling surfaces in \cite{Hung:2011nu} where the authors employed a conformal map from a hyperbolic cylinder to the causal evolution of a subregion, which is enclosed by a spherical entangling surface in flat space. The entanglement entropy of a spherical region in a conformal field theory (CFT) on a flat space may thus be expressed in terms of an integral of the thermal entropy of a CFT on a hyperbolic cylinder. This translates to the computation of the horizon entropy for a certain topological black hole through the Wald formula in the $AdS/CFT$ framework. The holographic R\'{e}nyi entropy $S^{(n)}(A)$ for a subsystem $A$ with spherical entangling surface in a $d$-dimensional conformal field theory ($CFT_d$) on a flat space is then given by
\begin{equation}
\label{ehresph}
S^{(n)}(A)=\mathcal{X}^{(n)}_d\,S(A)\,,
\end{equation}
\begin{equation}
\mathcal{X}^{(n)}_d=\frac{n}{2(n-1)}\left(2-x_n^{d-2}\left(1+x_n^2\right)\right),\qquad x_n=\frac{1}{nd}\left(1+\sqrt{1-2dn^2+d^2n^2}\right).\nonumber
\end{equation}
Note that the coordinate volume of the $(d-1)$ dimensional hyperbolic plane was computed in \cite{Hung:2011nu} to evaluate the entanglement entropy $S(A)$ in eq.\@ \eqref{ehresph}. The holographic R\'{e}nyi entropy of order half, which is related to the holographic entanglement negativity of a single subsystem, may then be obtained from eq.\@ \eqref{ehresph} as
\begin{equation}
\label{ehrehalf}
S^{(1/2)}(A)=\mathcal{X}_d\,S(A)\,,
\end{equation}
\begin{equation}
\mathcal{X}_d=\left(\frac{1}{2}x_d^{d-2}\left(1+x_d^2\right)-1\right),\qquad x_d=\frac{2}{d}\left(1+\sqrt{1-\frac{d}{2}+\frac{d^2}{4}}\right).\nonumber
\end{equation}
In eq.\@ \eqref{ehrehalf} and the rest of this appendix, we focus on the $n=1/2$ case and thus we denote $\mathcal{X}_d^{(1/2)}$ by $\mathcal{X}_d$. We note here that for the $AdS_3/CFT_2$ scenario, eq.\@ \eqref{ehrehalf} leads to $\mathcal{X}_2=3/2$, which is also the numerical factor obtained through the replica technique for a single interval.

Observe that the result in eq.\@ \eqref{ehrehalf} is valid only for a single connected subsystem with spherical geometry in the $AdS_{d+1}/CFT_d$ framework (which reduces to a single spatial interval in the $AdS_3/CFT_2$ scenario). For configurations involving multiple disjoint subsystems, the corresponding bulk geometry incorporates multiple disjoint branes whose backreaction on each other complicates the computation of the R\'{e}nyi entanglement entropy \cite{Dong:2016fnf}. However in the context of the $AdS_3/CFT_2$ framework involving two disjoint intervals in proximity (cross ratio of the intervals $x\to 1$) the above result continues to be valid. For this specific configuration, the following function involving integral values of the replica index $n$ vanishes in the limits $x\to 0$ and $x\to 1$ when $n>2$ as shown in \cite{Faulkner:2013yia}
\begin{equation}
\label{ejn}
J_n=I^{(n)}(A:B)-\frac{1}{2}\left(1+\frac{1}{n}\right)I(A:B)\,.
\end{equation}
In eq.\@ \eqref{ejn} $I^{(n)}(A:B)$ and $I(A:B)$ denote the R\'{e}nyi mutual information of order $n$ and the mutual information between the subsystems $A$ and $B$ respectively. These are defined as follows
\begin{equation}
\label{ermidef}
I^{(n)}(A:B)=S^{(n)}(A)+S^{(n)}(B)-S^{(n)}(AB)\,,
\end{equation}
\begin{equation}
\label{emidef}
I(A:B)=S(A)+S(B)-S(AB)\,.
\end{equation}
We may estimate the holographic R\'{e}nyi mutual information from eq.\@ \eqref{ejn} in the proximity limit $x\to 1$ as
\begin{equation}
\label{ermiapprx}
I^{(n)}(A:B)\approx\frac{1}{2}\left(1+\frac{1}{n}\right)I(A:B)\,.
\end{equation}
Eq.\@ \eqref{ermiapprx} may be continued analytically to $n\to 1/2$ as
\begin{equation}
\label{ermihalf}
I^{(1/2)}(A:B)\approx\frac{3}{2}I(A:B)\,.
\end{equation}
The condition in eq.\@ \eqref{ermihalf} requires that each of the R\'{e}nyi entanglement entropies of order half for the subsystems $A$, $B$ and $AB$, and their corresponding entanglement entropy in eqs.\@ \eqref{ermidef} and \eqref{emidef} are related through the prefactor $\mathcal{X}_2=3/2$ as shown in eq.\@ \eqref{ehrehalf}. We observe that the backreaction of the cosmic branes on each other remains non trivial if the cross ratio $x$ is not close to $0$ or $1$, and the relation in eq.\@ \eqref{ermihalf} may not hold. Thus our discussion of holographic entanglement negativity from now on will be restricted to the scenario when the intervals are in proximity ($x\to 1$) and the analytic continuation $n\to 1/2$ employed in eq.\@ \eqref{ermihalf} holds. Having discussed the holographic generalization of the R\'{e}nyi entanglement entropy and some of its properties for subsystems with spherical entangling surfaces, in section \ref{a2hen} we review the derivation of the holographic entanglement negativity.

\subsection{Holographic entanglement negativity}
\label{a2hen}

In this section we review the derivation of the holographic entanglement negativity of two disjoint subsystems in proximity in terms of the areas of the backreacting cosmic branes homologous to the subsystems as described in \cite{KumarBasak:2020ams}. This utilizes the holographic derivation of the entanglement negativity as shown in \cite{Dong:2021clv}. To this end we may define the R\'{e}nyi entanglement negativity of order $k$ following \cite{Calabrese:2012ew,Calabrese:2012nk} as
\begin{equation}
\label{erendef}
\mathcal{N}^{(k)}(A:B)=\mathrm{Tr}\left[\left(\rho_{AB}^{T_B}\right)^k\right]=\frac{Z\left[\mathcal{M}_k^{A,B}\right]}{\left(Z\left[\mathcal{M}\right]\right)^k}\,.
\end{equation}
Here $\mathcal{M}_k^{A,B}$ denotes the $k$-sheeted replicated manifold in which the individual copies of the subsystems $A$ and $B$ are respectively glued cyclically and anti-cyclically, $\mathcal{M}$ denotes the original manifold, and $Z$ denotes the path integral over the corresponding manifold. The entanglement negativity is then defined by the analytic continuation of even R\'{e}nyi negativities of order $k$ ($k=2n$) as given in eq.\@ \eqref{erendef} to $k\to 1$ ($n\to 1/2$) as
\begin{equation}
\label{eenren}
\mathcal{E}(A:B)=\lim_{n\to 1/2}\log\mathcal{N}^{(k=2n)}(A:B)\,.
\end{equation}
Note that the analytic continuation of the odd R\'{e}nyi negativities of order $k$ ($k=2n-1$) to $k\to 1$ ($n\to 1$) reproduces the trace normalization condition $\mathrm{Tr}\rho_{AB}^{T_B}=1$. Utilizing the AdS/CFT dictionary, we may express the ratio of the partition functions on the right hand side of eq.\@ \eqref{erendef} for even $k=2n$ at the leading order in terms of the corresponding gravitational on-shell actions in the bulk through the saddle point approximation as
\begin{equation}
\label{epibndblk}
\frac{Z\left[\mathcal{M}_{2n}^{A,B}\right]}{\left(Z\left[\mathcal{M}\right]\right)^{2n}}=\frac{Z\left[\mathcal{B}_{2n}^{A,B}\right]}{\left(Z\left[\mathcal{B}\right]\right)^{2n}}=\exp\left(2n\mathrm{I_{grav}}[\mathcal{B}]-\mathrm{I_{grav}}[\mathcal{B}_{2n}]\right)\,,
\end{equation}
where $\mathcal{B}_{2n}^{A,B}$ and $\mathcal{B}$ represent the bulk geometries corresponding to the manifolds $\mathcal{M}_{2n}^{A,B}$ and $\mathcal{M}$ respectively. Remarkably it was shown in \cite{Dong:2021clv} that unlike the R\'{e}nyi negativities, the replica symmetric gravitational saddle is the same for both even and odd $k$, and the corresponding entanglement negativity vanishes
\begin{equation}
\label{eenrsgs}
\mathcal{E}^{(\mathrm{sym})}(A:B)=\lim_{k\to 1}\log\mathcal{N}_{(k)}^{(\mathrm{sym})}=0.
\end{equation}
Quite interestingly the authors demonstrated that the bulk replica symmetry is broken by the dominant saddle corresponding to the entanglement negativity.

We now briefly describe the holographic construction of the replica non symmetric saddle involving $2n$ copies of the bulk manifold. To this end we consider three codimension one homology hypersurfaces $\Sigma_A$, $\Sigma_B$ and $\Sigma_{\overline{AB}}$ along which each such manifold has cuts. These hypersurfaces are non-overlapping and obey the homology conditions $\partial\Sigma_X=X\cup\gamma_X$, $\gamma_X$ being the codimension two bulk hypersurface homologous to a subsystem $X$ ($X=A,B,\overline{AB}$). These hypersurfaces are then sewed as follows.
\begin{itemize}
\item
$\Sigma_A$: odd copies of bulk manifold are sewed cyclically and even copies are sewed to themselves,
\item
$\Sigma_B$: even copies of bulk manifold are sewed anti-cyclically and odd copies are sewed to themselves,
\item
$\Sigma_{\overline{AB}}$: all copies are sewed pairwise.
\end{itemize}
We note that whereas the replica symmetry holds for the above construction in the boundary, it is explicitly broken while going from $\mathbb{Z}_{2n}\to \mathbb{Z}_n$ in the bulk. This observation motivated the authors in \cite{Dong:2021clv} to consider instead the $\mathbb{Z}_n$ quotient of the original manifold given as
\begin{equation}
\label{eznquot}
\hat{\mathcal{B}}_{2n}^{A,B(\mathrm{nsym)}}=\mathcal{B}_{2n}^{A,B(\mathrm{nsym)}}/\mathbb{Z}_n\,.
\end{equation}
The quotienting described in eq.\@ \eqref{eznquot} modifies the corresponding bulk gravitational on-shell actions as
\begin{equation}
\label{equotact}
\mathrm{I_{grav}}\left[\mathcal{B}_{2n}^{A,B(\mathrm{nsym)}}\right]\equiv n\mathrm{I_{grav}}\left[\hat{\mathcal{B}}_2^{A,B(\mathrm{nsym)}}\right]=n\mathrm{I_{grav}}\left[\mathcal{M}_2^{AB},\gamma_{A_1}^{(n)},\gamma_{B_2}^{(n)}\right].
\end{equation}
Here $\mathrm{I_{grav}}\left[\mathcal{M}_2^{AB},\gamma_{A_1}^{(n)},\gamma_{B_2}^{(n)}\right]$ is the on-shell bulk action for the alternative configuration of the quotiented bulk manifold $\hat{\mathcal{B}}_2^{A,B(\mathrm{nsym)}}$, where it has conical defects along codimension two surfaces $\gamma_{A_1}^{(n)}$ and $\gamma_{B_2}^{(n)}$, and $\mathcal{M}_2^{AB}$ as its asymptotic boundary. Subscripts 1 and 2 denote the copy where the corresponding surface resides. We are now in a position to obtain an expression for the R\'{e}nyi negativity by substituting eq.\@ \eqref{equotact} into eq.\@ \eqref{epibndblk} and inserting the result into eq.\@ \eqref{erendef} as
\begin{equation}
\label{erenblkact}
\log\mathcal{N}_{(2n)}^{(\mathrm{even,nsym})}=-n\left(\mathrm{I_{grav}}\left[\mathcal{M}_2^{AB},\gamma_{A_1}^{(n)},\gamma_{B_2}^{(n)}\right]-2\mathrm{I_{grav}}\left[\mathcal{B}\right]\right).
\end{equation}
Note that eq.\@ \eqref{erenblkact} was derived in \cite{Dong:2021clv} on the basis of the assumption that the system comprising the subsystems $A$, $B$ and $\overline{AB}$ is in a tripartite pure state. A result for the bulk action away from $n=1$ described in \cite{Nakaguchi:2016zqi,Kawabata:2021vyo} was employed in \cite{KumarBasak:2020ams} to express the on-shell bulk action as
\begin{equation}
\label{eblkactarea}
\mathrm{I_{grav}}\left[\mathcal{M}_2^{AB},\gamma_{A_1}^{(n)},\gamma_{B_2}^{(n)}\right]=2\mathrm{I_{grav}}\left[\mathcal{B}\right]+\frac{\mathcal{A}^{(1/2)}\left(\gamma_{AB}\right)}{4G_N}+\left(1-\frac{1}{n}\right)\frac{\mathcal{A}^{(n)}\left(\gamma_A\right)+\mathcal{A}^{(n)}\left(\gamma_B\right)}{4G_N}\,.
\end{equation}
Substituting eq.\@ \eqref{eblkactarea} into eq.\@ \eqref{erenblkact} and inserting the result further into eq.\@ \eqref{eenren} we arrive at the following expression for the holographic entanglement negativity
\begin{align}
\mathcal{E}(A:B)&=\frac{1}{8G_N}\left[\mathcal{A}^{(1/2)}\left(\gamma_A\right)+\mathcal{A}^{(1/2)}\left(\gamma_B\right)-\mathcal{A}^{(1/2)}\left(\gamma_{AB}\right)\right]\nonumber\\\nonumber\\&=\frac{1}{2}\left[S^{(1/2)}(A)+S^{(1/2)}(B)-S^{(1/2)}(AB)\right]\nonumber\\\nonumber\\&=\frac{1}{2}I^{(1/2)}(A:B)\,.\label{ehenhmi}
\end{align}
Starting from eq.\@ \eqref{ehenhmi}, in section \ref{a2hensg} we follow \cite{KumarBasak:2020ams} to describe a plausible proof of the holographic construction for the entanglement negativity of a single interval as advanced in \cite{Chaturvedi:2016rcn}.

\subsection{Holographic entanglement negativity for a single interval}
\label{a2hensg}

In this section we describe a plausible derivation of the holographic proposal for the entanglement negativity of a single interval in the context of $AdS_3/CFT_2$, following the work in \cite{KumarBasak:2020ams}. To this end we establish an important relation below for the R\'{e}nyi mutual information of order half. Consider a tripartite system $ABC$ (consisting of the subsystems $A$, $B$ and $C$) in a pure state. We have
\begin{align}
&I^{(1/2)}(A:BC)-I^{(1/2)}(A:C)\nonumber\\\nonumber\\&=\left[S^{(1/2)}(A)+S^{(1/2)}(BC)-S^{(1/2)}(ABC)\right]-\left[S^{(1/2)}(A)+S^{(1/2)}(C)-S^{(1/2)}(AC)\right]\nonumber\\\nonumber\\&=S^{(1/2)}(BC)+S^{(1/2)}(AC)-S^{(1/2)}(C)-S^{(1/2)}(ABC)\,.\label{ermiree}
\end{align}
Since the R\'{e}nyi entropy of a system in a pure state is zero, we have $S^{(1/2)}(ABC)=0$. Furthermore, for a bipartite system in a pure state, the entanglement negativity is given by the R\'{e}nyi entropy of order half, $\mathcal{E}(A:BC)=S^{(1/2)}(A)$ and $\mathcal{E}(BC:A)=S^{(1/2)}(BC)$. Since $\mathcal{E}(A:BC)=\mathcal{E}(BC:A)$, we have $S^{(1/2)}(A)=S^{(1/2)}(BC)$. Relabeling $A$, $B$ and $C$, we further obtain $S^{(1/2)}(B)=S^{(1/2)}(AC)$ and $S^{(1/2)}(C)=S^{(1/2)}(AB)$. These relations may be organized as follows
\begin{equation}
\label{ereerel}
\begin{aligned}
&S^{(1/2)}(BC)=S^{(1/2)}(A)\,,\qquad\qquad&&S^{(1/2)}(AC)=S^{(1/2)}(B)\,,\\&S^{(1/2)}(C)=S^{(1/2)}(AB)\,, &&S^{(1/2)}(ABC)=0\,.
\end{aligned}
\end{equation}
Substituting the relations in eq.\@ \eqref{ereerel} into eq.\@ \eqref{ermiree} we obtain $I^{(1/2)}(A:BC)-I^{(1/2)}(A:C)=I^{(1/2)}(A:B)$, which may be re-expressed as
\begin{equation}
\label{ermirel}
I^{(1/2)}(A:BC)=I^{(1/2)}(A:B)+I^{(1/2)}(A:C)\,.
\end{equation}
Having derived the relation in eq.\@ \eqref{ermirel}, we are now in a position to show that the holographic construction in \cite{Chaturvedi:2016rcn} may be deduced from eq.\@ \eqref{ehenhmi}. The corresponding configuration may be briefly described as an interval $A$ with two auxiliary intervals $B_1$ and $B_2$ on either side. In the bipartite limit $B_1B_2\to A^c$, the system $AB_1B_2$ is in a pure state, and we may apply the result in eq.\@ \eqref{ermirel} to this system to obtain the following relation
\begin{equation}
\label{ermirelsg}
\lim_{B_1B_2\to A^c}I^{(1/2)}(A:B_1B_2)=\lim_{B_1B_2\to A^c}\left[I^{(1/2)}(A:B_1)+I^{(1/2)}(A:B_2)\right]\,.
\end{equation}
The holographic entanglement negativity for this configuration may then be computed by substituting eq.\@ \eqref{ermirelsg} into eq.\@ \eqref{ehenhmi} as
\begin{equation}
\label{ehenrmisg}
\mathcal{E}=\lim_{B_1B_2\to A^c}\left.\frac{1}{2}\middle[I^{(1/2)}(A:B_1)+I^{(1/2)}(A:B_2)\right]\,.
\end{equation}
Through the definition of R\'{e}nyi mutual information, eq.\@ \eqref{ehenrmisg} may be expressed as
\begin{equation}
\label{ehenreesg}
\mathcal{E}=\lim_{B_1B_2\to A^c}\left.\frac{1}{2}\middle[2S^{(1/2)}(A)+S^{(1/2)}(B_1)+S^{(1/2)}(B_2)-S^{(1/2)}(A\cup B_1)-S^{(1/2)}(A\cup B_2)\right].
\end{equation}
Since each of the intervals involved on the right hand side of eq.\@ \eqref{ehenreesg} is a single connected subsystem with spherical geometry, eq.\@ \eqref{ehrehalf} applies to each of them, and we have $S^{(1/2)}(X)=\mathcal{X}_2\,S(X)=\frac{3}{2}S(X)$, where $X$ represents each of the intervals. We may now re-express eq.\@ \eqref{ehenreesg} as
\begin{equation}
\label{ehenheesg}
\mathcal{E}=\lim_{B_1B_2\to A^c}\left.\frac{3}{4}\middle[2S(A)+S(B_1)+S(B_2)-S(A\cup B_1)-S(A\cup B_2)\right].
\end{equation}
Utilizing the RT formula $S(X)=\mathcal{L}_X/4G_N$ and eq.\@ \eqref{ehenheesg}, we may finally arrive at the following expression for the holographic entanglement negativity for a single interval in the $AdS_3/CFT_2$ framework
\begin{equation}
\label{ehenlensg}
\mathcal{E}=\lim_{B\to A^c}\left.\frac{3}{16G_N}\middle[2\mathcal{L}_A+\mathcal{L}_{B_1}+\mathcal{L}_{B_2}-\mathcal{L}_{A\cup B_1}-\mathcal{L}_{A\cup B_2}\right],
\end{equation}
where $B\equiv B_1\cup B_2$. Eq.\@ \eqref{ehenlensg} may be readily identified as the holographic conjecture for the entanglement negativity of a single interval in the $AdS_3/CFT_2$ scenario as proposed in \cite{Chaturvedi:2016rcn}. This concludes the review of the derivation of the holographic proposal for the entanglement negativity of a single interval.

\section{Non extremal and extremal $AdS_{d+1}$-RN}
\label{arnhd}

The numerical constants $b_0$ and $b_1$ appearing in eqs.\@ \eqref{eephd} and \eqref{eteffhd} are given as follows
\begin{equation}
\label{eb0}
b_0=\frac{2(d-1)}{d-2}\;,
\end{equation}
\\
\begin{equation}
\label{eb1}
b_1=\frac{d}{d-2}\;.
\end{equation}
\\

The constants $\mathcal{S}_0$ and $\mathcal{S}_1$ appearing in eqs.\@ \eqref{eareahdnescpltst}, \eqref{eareahdnescpltbt} and \eqref{eareahdexscp} are given as follows
\begin{equation}
\label{eS0}
\mathcal{S}_0=\frac{2^{d-2}\pi^{\frac{d-1}{2}}\Gamma\left[-\frac{d-2}{2(d-1)}\right]}{(d-1)\Gamma\left[\frac{1}{2(d-1)}\right]}\left(\frac{\Gamma\left[\frac{d}{2(d-1)}\right]}{\Gamma\left[\frac{1}{2(d-1)}\right]}\right)^{d-2},
\end{equation}
\\
\begin{equation}
\label{eS1}
\mathcal{S}_1=\frac{\Gamma\left[\frac{1}{2(d-1)}\right]^{d+1}2^{-d-1}\pi^{-\frac{d}{2}}}{\Gamma\left[\frac{d}{2(d-1)}\right]^d\Gamma\left[\frac{1}{2}+\frac{1}{d-1}\right]}\left(\frac{\Gamma\left[\frac{1}{d-1}\right]}{\Gamma\left[-\frac{d-2}{2(d-1)}\right]}+\frac{2^{\frac{1}{d-1}}(d-2)\Gamma\left[1+\frac{1}{2(d-1)}\right]}{\sqrt{\pi}(d+1)}\right).
\end{equation}
\\

The numerical constants $N_0$ and $N_1$ appearing in eq.\@ \eqref{eareaauxhdnescpltst} are given as follows
\begin{equation}
\label{eN0}
\begin{aligned}
N_0&=2\left(\frac{\sqrt{\pi}~\Gamma\left[-\frac{d-2}{2(d-1)}\right]}{2(d-1)\Gamma\left[\frac{1}{2(d-1)}\right]}\right)\\\\&+2\int_0^1dx\left(\frac{\sqrt{1-x^{2(d-1)}}}{x^{d-1}\sqrt{1-b_0x^d+b_1x^{2(d-1)}}}-\frac{1}{x^{d-1}\sqrt{1-x^{2(d-1)}}}\right),
\end{aligned}
\end{equation}
\\
\begin{equation}
\label{eN1}
N_1=\int_0^1dx\left(\frac{x\sqrt{1-x^{2(d-1)}}}{\sqrt{1-b_0x^d+b_1x^{2(d-1)}}}\right)\left(\frac{1-x^{d-2}}{1-b_0x^d+b_1x^{2(d-1)}}\right).
\end{equation}
\\

The function $\gamma_d\left(\frac{\mu }{T}\right)$ appearing in eq.\@ \eqref{eareaauxhdnescpltbt} is given as follows
\begin{equation}
\label{gamma}
\gamma_d\left(\frac{\mu }{T}\right)=N(1)+\frac{d^2(d-2)}{16\pi^2(d-1)}\left(\frac{\mu }{T}\right)^2\int_0^1dx\left(\frac{x\sqrt{1-x^{2(d-1)}}}{\sqrt{1-x^d}}\right)\left(\frac{1-x^{d-2}}{1-x^d}\right)+\mathcal{O}\left[\left(\frac{\mu }{T}\right)^4\right],
\end{equation}
where the expression for the numerical constant $N(\varepsilon)$ is given in eq.\@ \eqref{eNep}.

The numerical constant $N(\varepsilon)$ appearing in eqs.\@ \eqref{eareaauxhdexscp} and \eqref{gamma} is given as follows
\begin{equation}
\label{eNep}
N(\varepsilon)=2\left(\frac{\sqrt{\pi}\Gamma\left[-\frac{d-2}{2(d-1)}\right]}{2(d-1)\Gamma\left[\frac{1}{2(d-1)}\right]}\right)+2\int_0^1dx\left(\frac{\sqrt{1-x^{2(d-1)}}}{x^{d-1}\sqrt{f(z_Hx)}}-\frac{1}{x^{d-1}\sqrt{1-x^{2(d-1)}}}\right).
\end{equation}

\end{appendices}

\bibliographystyle{JHEP}
\bibliography{HensgRNRef}

\end{document}